\documentclass[12pt]{article}
\usepackage[dvips]{graphicx}
\usepackage{epsfig,amsmath}
\begin{document}
\thispagestyle{empty}
\begin{center}

{\Large\bf{How is transversity related to helicity\\

\vskip 0.5cm
for quarks and antiquarks inside the proton?}}

\vskip1.4cm
{\bf Claude Bourrely}
\vskip 0.3cm
D\'epartement de Physique, Facult\'e des Sciences de Luminy,\\
Universit\'e de la M\'editerran\'ee - Aix-Marseille II,\\
13288 Marseille, Cedex 09, France\\
\vskip 0.5cm
{\bf Franco Buccella}
\vskip 0.3cm
Dipartimento di Scienze Fisiche, Universit\`a di Napoli,\\
Via Cintia, I-80126, Napoli
and INFN, Sezione di Napoli, Italy
\vskip 0.5cm
{\bf Jacques Soffer}
\vskip 0.3cm
Physics Department, Temple University\\
Barton Hall, 1900 N, 13th Street\\
Philadelphia, PA 19122-6082, USA
\vskip 1.5cm
{\bf Abstract}\end{center}
We consider the quark and antiquark transversity distributions inside a
polarized proton and we study how they are expected to be
related to the corresponding helicity distributions, both in sign and
magnitude. Our considerations lead to simple predictions in good agreement
with their first determination for light quarks from experimental data. We
also give our predictions for the light antiquarks transversity distributions,
so far unknown.

\vskip 0.5cm


\vskip 0.5cm

\noindent UNIV. NAPLES DSF 3-2009
\newpage
%%%%%%%%%%%%%%%%%%%%%%%%%%%%%%%%%%%%%%%%%%%%%%%%%%%%%%%%%%%%%%%%%%%%%%%
Our understanding of the proton spin structure has greatly improved over
the last twenty years or so due, on the one hand, to significant progress on
the theoretical
side and on the other hand, to several dedicated experiments at CERN, DESY,
JLab and SLAC on polarized
deep inelastic scattering (DIS). More recently, the advent of the
polarized $pp$ collider at RHIC-BNL has opened up a new era for a better
knowledge of the proton spin
structure and also for testing the spin sector of perturbative QCD. \\
The main source of information on the internal proton structure lies in the
parton
distributions. If ${\cal A}(x)$ denotes the quark distribution in a proton, as
a density matrix in both
the quark and proton spin, it will be expressed in terms of direct products of
two
Pauli matrices $\sigma_i $ and the unit matrix $I$. Then, by choosing the
$z$-axis along the proton
momentum, the $x$-axis and $y$-axis normal to it, ${\cal A}(x)$ reads
\begin{equation}
{\cal A}(x) = q(x)I\otimes I - \Delta q(x)\sigma_z\otimes\sigma_z - \delta
q(x)(\sigma_x\otimes\sigma_y +
\sigma_y\otimes\sigma_x)~,
\label{def}
\end{equation}
where $q(x)$ is the {\it unpolarized} distribution, whereas for the {\it
polarized}
distributions, one must distinguish helicity distributions inside a
longitudinally polarized proton,
denoted for quarks by $\Delta q(x)$ and transversity distributions inside a
transversely polarized proton, denoted by $\delta q(x)$. These last two
distributions are required for a complete description of the quark spin
in the proton at leading twist.

The vast programme of unpolarized DIS data taking at HERA has led to a rather
precise determination of the quark ($q$), antiquark ($\bar q$) and gluon ($G$)
unpolarized
distributions, which is very relevant to study hadronic processes at LHC-CERN
\cite{ditt}.  The helicity distributions have been determined so far, with a
reasonable precision level, because they can be directly extracted from
polarized DIS but this is not the case for the transversity distributions,
which are not easily accessible, being chiral odd, they decouple from DIS
\cite{bdr}. However, there is a strong bound on transversity, resulting from
positivity and derived a few years ago \cite{js}; it involves helicity and
reads

\begin{equation}
q(x) + \Delta q(x) \geq 2|\delta q(x)|~,
\label{bound}
\end{equation}
for quarks and similarly for antiquarks, which is obviously more severe for
negative quark helicity distributions.\\
Let us first examine some arguments to appreciate the relevance of this
positivity constraint \footnote{For a recent review on positivity constraints
for spin observables see Ref.~\cite{aerst}.}. In
the non-relativistic limit, transversity and helicity coincide indeed, that is
\begin{equation}
 \delta q(x) = \Delta q(x)~,
\label{equality}
\end{equation}
so the bound is trivially fullfilled, provided $\Delta q(x) \geq 0$. On the
contrary if $\Delta q(x) \leq 0$, the
bound implies
\begin{equation}
\Delta q(x)/q(x) \geq -1/3~.
\label{neg}
\end{equation}
This simple remark is stressing the importance of the sign of $\Delta q(x)$ and
we now turn to review what is known about this sign, for the different flavor
quarks and antiquarks.
Concerning the light quarks helicity distibutions, 
it is well established that,
$\Delta u(x)>0$ and $\Delta d(x)<0$, and according to some very accurate JLab
data \cite{JLab}, the ratio
$\Delta d(x)/d(x)$ is close to -1/3 at $x=0.6$, with a possible violation of
Eq.~(\ref{equality}) \footnote{It is interesting
to note that in Ref.~\cite{jf}, one argues that $0 \geq \Delta d(x)/d(x) \geq
-1/3$ for all $x$,  based on general properties of a three quark bound state
obeying the Pauli principle.}, which leads to a
contradiction with the assumption Eq.~(\ref{equality}). This trend has been
correctly predicted by our quantum statistical approach for unpolarized and
polarized parton
distributions \cite{bbs1}, where the non-diffractive part, the only one
contributing to helicity distributions, is
given by Fermi-Dirac functions, namely the first term in the r.h.s of Eq.~(14)
of \cite{bbs1} for quarks and the first term of the r.h.s of Eq.~(15) for
antiquarks. The parton distributions determined in \cite{bbs1} have been
successfully compared with new experimental results in \cite{bbs2,bbs3}.
Concerning the light antiquark helicity distributions, the statistical 
approach
imposes a strong relationship to the corresponding quark helicity
distributions. In particular, it predicts $\Delta \bar u(x)>0$ and $\Delta \bar
d(x)<0$, with almost the same magnitude, in contrast with the
simplifying assumption $\Delta \bar u(x)=\Delta \bar d(x)$, often adopted in
the literature. The COMPASS experiment
at CERN has measured the valence quark helicity distributions, defined as
$\Delta q_v(x)= \Delta q(x)-\Delta \bar q(x)$. These recent results displayed
in Fig.~1 are compared to our prediction and the data give
$\Delta \bar u(x) + \Delta \bar d(x) \simeq 0$, which implies either small or
opposite values for $\Delta \bar u(x)$ and $\Delta \bar d(x)$. Indeed $\Delta
\bar u(x)>0$ and $\Delta \bar d(x)<0$ are predicted both by the
chiral quark soliton model (CQSM) \cite{cqsm}-\cite{dress00} and 
the statistical approach \cite{bbs1} and it leads to
a positive contribution of the sea to the Bjorken sum rule \cite{bj}.\\
Although strange quarks and antiquarks $s$ and $\bar s$ play a fundamental
role in the nucleon structure, they are much less known than the parton
distributions for the light quarks $u$ and $d$. For completeness, let us just
mention that
we have extended the statistical approach to this case and we have found that
$\Delta s(x)$
and $\Delta \bar s(x)$ are both negative for all $x$ values \cite{bbs7}.
Negative values are also predicted
by the flavor SU(3) CQSM \cite{WA03}, with the magnitude of $\Delta \bar s(x)$
much smaller than that of
$\Delta s(x)$. The conclusion on recent COMPASS data \cite{rw} is that $\Delta
s(x)$ is nearly zero or
slightly negative depending on the choice of the kaon fragmentation functions.
\\
So to summarize this discussion about the sign of the quark and antiquark
helicity distributions, it seems clear
that $\Delta u(x)>0$ and $\Delta d(x)<0$, whereas for the remaining ones the
sign is not yet firmly established
which requires a more elaborate future experimental investigation.\\
Let us now recall that, by studying the effects due to the Melosh-Wigner
rotation \cite{ss,mss}, an {\it approximate} relation between transversity and
helicity distributions was derived, namely
\begin{equation}
\Delta q_{RF}(x) + \Delta q(x) = 2\delta q(x)~,
\label{melosh}
\end{equation}
where $\Delta q_{RF}(x)$ measures the quark constituent spin in the proton rest
frame, which differs from $\Delta q(x)$ due to the Melosh-Wigner rotation. It looks
similar to Eq.~(\ref{bound}) and it is compatible with it, when $\Delta q(x)
\geq 0$, since
$q(x) \geq \Delta q_{RF}(x)$. One can make another interesting remark to stress
the difference between
$u$-quarks and $d$-quarks. By considering the first moments, according to
SU(6), one has
\begin{equation}
\Delta u_{RF}=4/3~~~~~~ \mbox{and}~~~~~~\Delta d_{RF}=-1/3~,
\label{su(6)}
\end{equation}
whereas the first moments at $Q^2=0$ are
\begin{equation}
\Delta u=2F~~~~~~ \mbox{and}~~~~~~\Delta d=F-D~,
\label{FD}
\end{equation}
where $F$ and $D$ are the two parameters introduced by Cabibbo in his famous
work on the weak current
of the hadrons \cite{nc}, whose values are $F=0.464$ and $D=0.787$. Therefore
\begin{equation}
\Delta u < \Delta u_{RF}~~~~~~ \mbox{and}~~~~~~\Delta d \simeq \Delta d_{RF}~.
\label{RF}
\end{equation}
This is reasonable, since the Melosh-Wigner rotation depends on the transverse
momentum of the
quarks, which is expected, in the framework of the quantum statistical
approach, to be larger for the $u$-quarks than for the $d$-quarks. This is due
to the fact that the $u$-quark distribution, which has the largest first
moment, occupies broader regions of the phase-space both in the longitudinal
and transverse directions \cite{bbs4}.\\
So by taking $\Delta d(x) \simeq \Delta d_{RF}(x)$ and by using
Eq.~(\ref{melosh}), for the $d$-quark, which gives
$\Delta d(x) \simeq \delta d(x)$, we see that
the positivity bound Eq.~(\ref{bound}) implies again $\Delta d(x)/d(x)\geq
-1/3$. So we have some evidence that the assumption Eq.~(\ref{equality})
combined with the Jlab data \cite{JLab} might lead to a violation of the
positivity bound for the $d$-quark and we propose to assume instead
\begin{equation}
 \delta q(x) = \kappa\Delta q(x)~,
\label{prop}
\end{equation}
where $\kappa$ is a normalization factor which will be taken at the largest
value, in such a way that the positivity bound is satisfied for the $d$-quark.
By using the statistical distributions \cite{bbs3}, we show in Fig.~2 the
resulting quark transversity distributions for $Q^2=2.4 \mbox{GeV}^2$, where we
took $\kappa=0.6$, for $u$ and $d$ flavors, together with the positivity
bounds. Although this value of $\kappa$ was chosen to satisfy positivity for
the $d$-quark, for simplicity, the same value was kept for the $u$-quark. In
Ref.~\cite{ma}, the first extraction of the $u$- and $d$-quark transversity
distributions had been obtained, by a combined global analysis of the 
azimuthal
asymmetries in semi-inclusive polarized DIS measured by HERMES at DESY and
COMPASS at CERN and those in $e^+e^- \to h_1h_2X$ unpolarized processes by the
Belle Collaboration at KEK. The agreement between these experimental results
\cite{ma} and the curves displayed in Fig.~2 is rather satisfactory, within the
uncertainties. One should note that since $\kappa < 1$, at least at low $Q^2$,
transversity is {\it smaller} than helicity, in agreement
with Ref.~\cite{ma}, but in contrast
with the results of Ref.~\cite{mss}, which was predicting transversity {\it
larger} than helicity. Moreover, in
Ref.~\cite{mss} the signs of the quark transversity distributions is correctly
predicted, but the magnitudes are far too large. A comparative analysis of
transversity and helicity was also presented in a more recent work \cite{mw},
with some predictions from the CQSM, and although the $d$-quark transversity
distribution has the correct sign and magnitude, it is definitely too large 
for the corresponding $u$-quark.\\
Another relevant point, not yet mentioned so far, concerns the $Q^2$ evolution
because it is important
to take into account the different $Q^2$ evolution of the transversity and
helicity distributions as
discussed recently in Ref.~\cite{mw1}. In particular, one should remember that
the scale dependence
of the tensor charges is fairly strong in sharp contrast to the case of the
axial charges. This means that the simplifying assumption Eq.~(\ref{prop}) at
the initial scale, becomes an approximation at
higher scales, an obvious limitation for very accurate predictions, because
$\kappa$ should depend on $Q^2$.\\
For completeness, by using Eq.~(\ref{prop}) for antiquarks, we also give the
resulting transversity distributions displayed on Fig.~3, which have the same
signs as the quark transversity distributions and satisfy positivity. 
As direct
consequence of this, the double transverse spin asymmetry  $A_{TT}$, for
Drell-Yan muon pair production is expected to be positive, for $pp$ collisions
as well as for $\bar pp$ collisions. We show in Fig.~4 the expected
asymmetries at RHIC-BNL for $pp$ and at the new FAIR accelerator complex at
Darmstadt for $\bar pp$, where
there are speculations for producing a polarized $\bar p$ beam \cite{pax}. In
both cases the asymmetry increases with larger
$M$. Clearly we expect the corresponding double helicity asymmetry $A_{LL}$ to
be larger as already pointed out in
several other cases \cite{jsa}, a theoretical observation which must be
carefully checked by experiments. \\
Finally, it is interesting to remark that, at least at low $Q^2$, the quark
transversity and helicity
distributions have rather similar shapes, which was not necessarily
anticipated.
Our knowledge on the transversity distributions is at the earlier stage and we
look forward to precise data, to  improve this important aspect of the proton
spin structure.

%%%%%%%%%%%%%%%%%%%%%%%%%%%%%%%%%%%%%%%%%%%%%%%%%%%%%%%%%%%%%%%%%%

\newpage

\begin{figure}[htbp]
\begin{center}
\includegraphics[angle=-90,width=13.0cm]{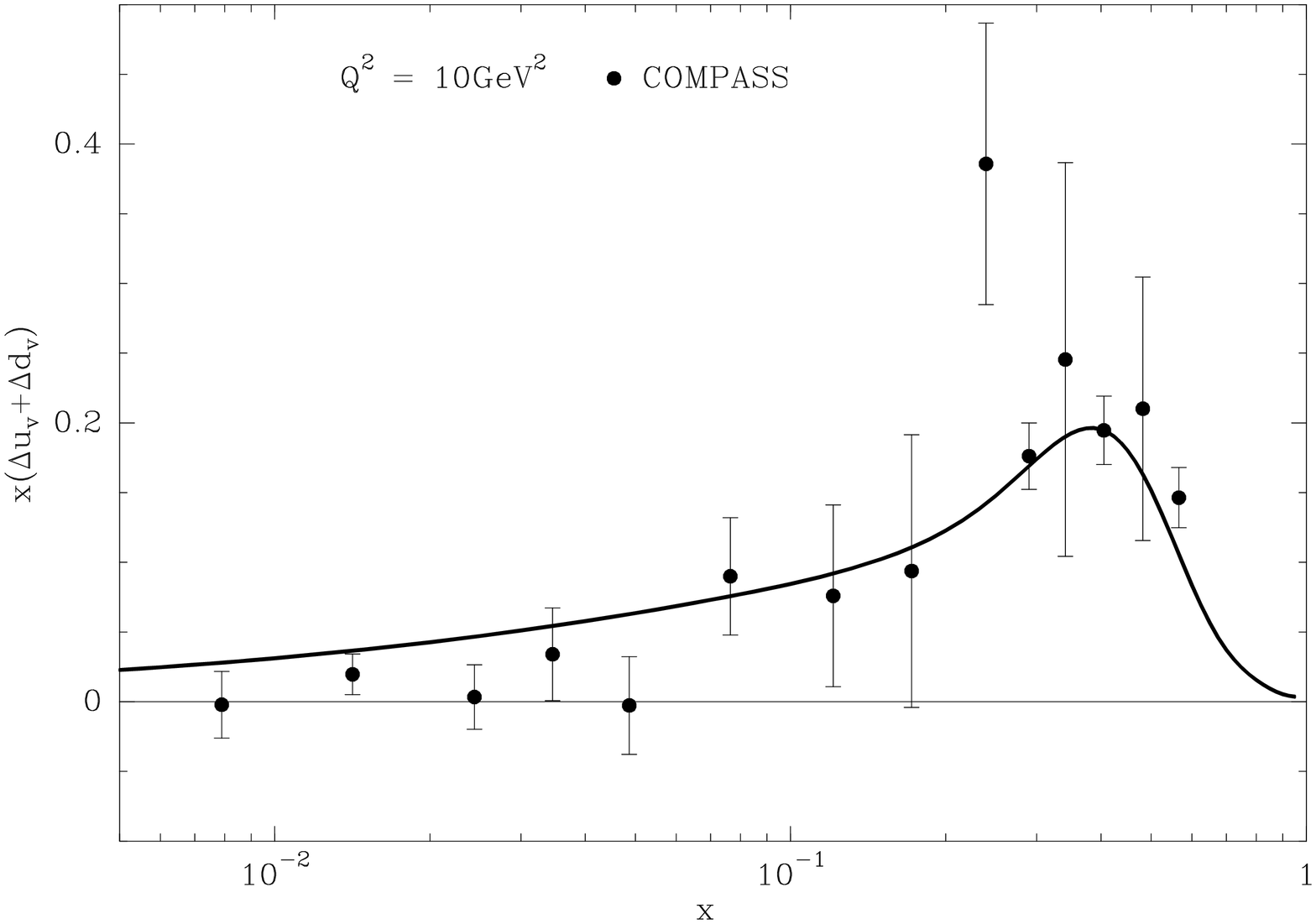}
\caption[*]{\baselineskip 1pt
The valence quark helicity distributions, versus $x$ and
evolved at $Q^2 = 10 \mbox{GeV}^2$. The solid curve is the prediction of the
statistical approach and the data points come from Ref.~\cite{compass}.}
\end{center}
\label{fi:sumudv09}
\end{figure}

\begin{figure}[htb]
\begin{center}
\leavevmode {\epsfysize=15.cm \epsffile{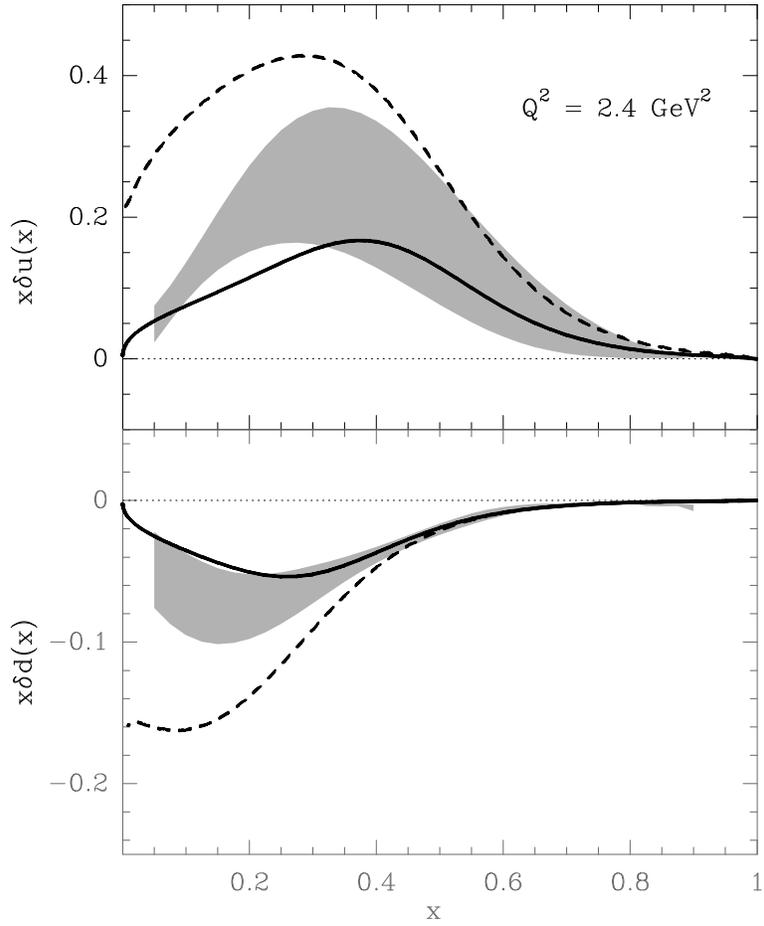}}
\end{center}
  \vspace*{-5mm}
\caption[*]{\baselineskip 1pt
The resulting quark transversity distributions for $u$ and $d$ flavors, as
a function of $x$ for  $Q^2 = 2.4  \mbox{GeV}^2$. The dashed
lines are the positivity bounds and the shaded areas are the uncertainties
bands
obtained in Ref.~\cite{ma}.}
\label{fi:qversusx}
\vspace*{-1.0ex}
\end{figure}

\begin{figure}[htb]
\begin{center}
\leavevmode {\epsfysize=15.cm \epsffile{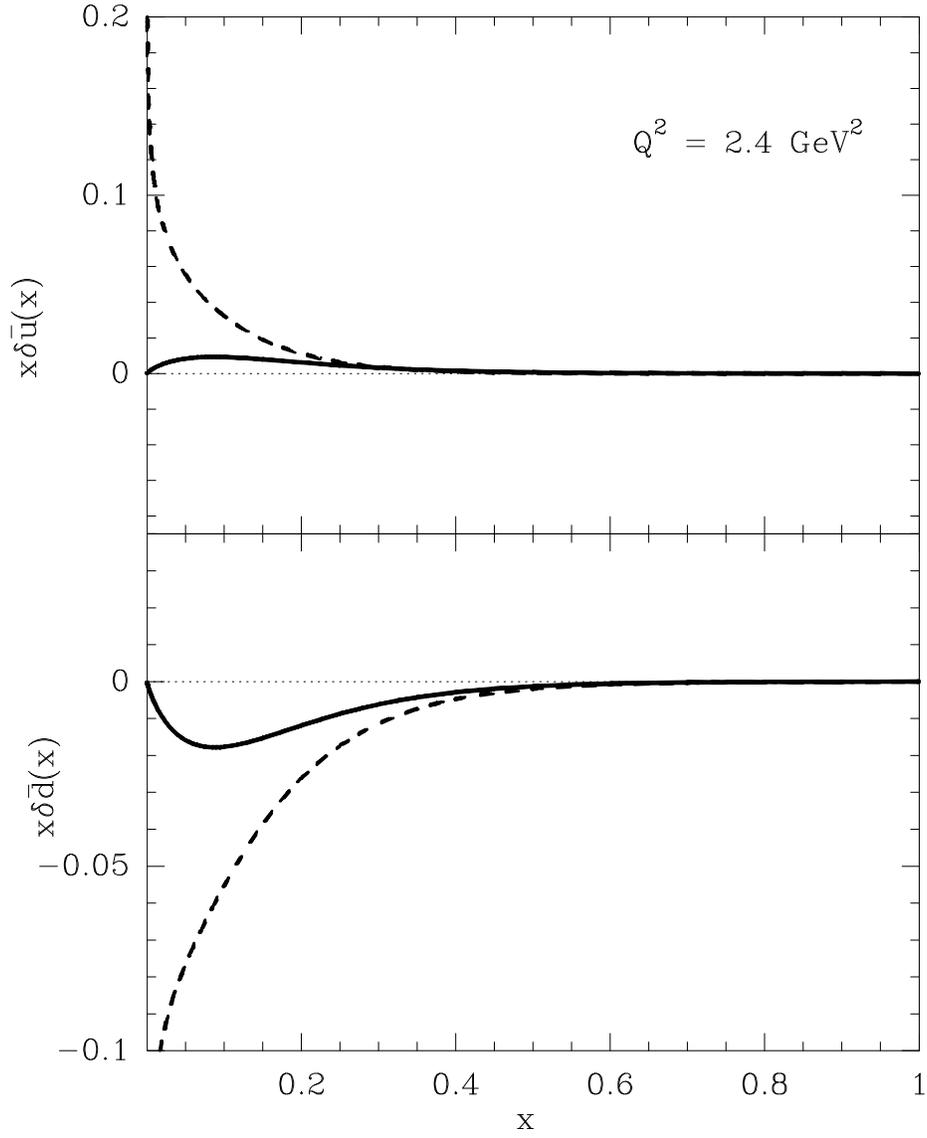}}
\end{center}
  \vspace*{-5mm}
\caption[*]{\baselineskip 1pt
The resulting antiquark transversity distributions for $u$ and $d$ flavors, as
a function of $x$ for  $Q^2 = 2.4  \mbox{GeV}^2$. The dashed
lines are the positivity bounds.}
\label{fi:barqversusx}
\vspace*{-1.0ex}
\end{figure}

\begin{figure}[htb]
\begin{center}
\leavevmode {\epsfysize=15.cm \epsffile{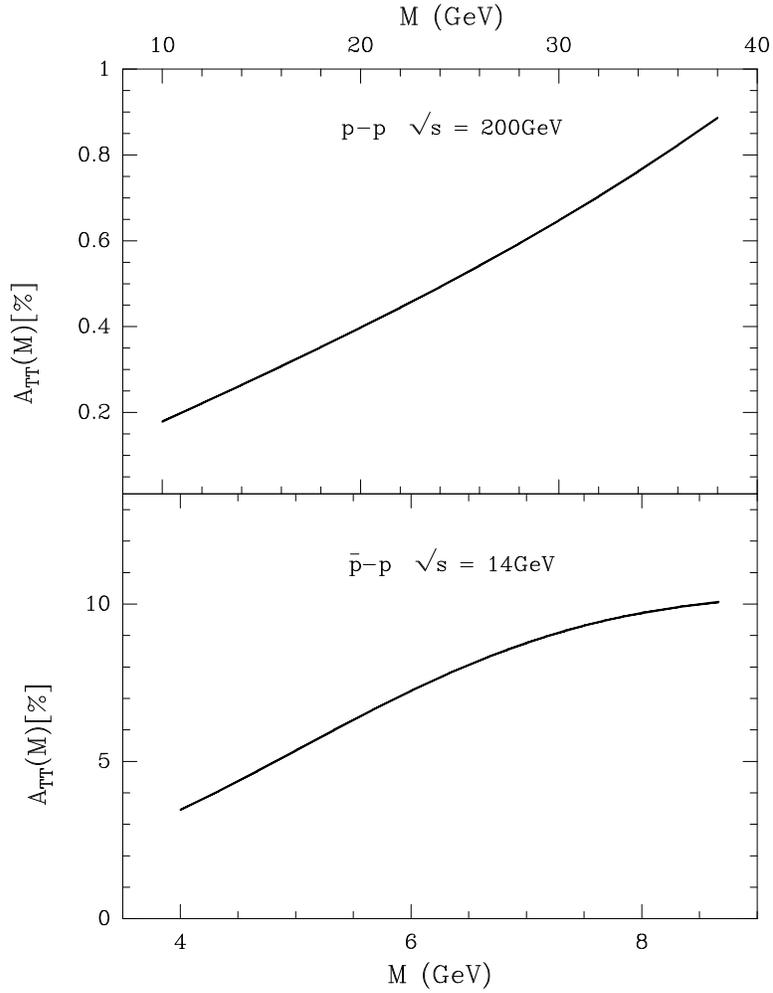}}
\end{center}
  \vspace*{-5mm}
\caption[*]{\baselineskip 1pt
The predicted double transverse spin asymmetry for Drell-Yan muon pair
production at zero rapidity, versus
the muon pair mass $M$. Upper part
for $pp$ collisions at RHIC-BNL. Lower part for $\bar p p$ collisions at FAIR
Darmstadt.}
\label{fi:att}
\vspace*{-1.0ex}
\end{figure}

\end{document}